# Manifestation of a General Coherent State Superposition without Nonlinear Effects

Gh. Asadi Cordshooli[1], Mehdi Mirzaee[1 *]

Department of Physics, Faculty of Science, Arak University, Arak 38156-8-8349, Iran

We report the formation of a general superposition of coherent states in exact analytical solution of the Schrodinger's equatin for atom-photon interaction by taking into account the role of virtual photons. Despite some known superposition of quantum states, the general superposition state introduced in this letter constructed without any nonlinear effect. The Yurke-Stoler state and its special cases, the even and odd cat states, obtained as some simple examples of the method. A general experimental setup is proposed to realize a variety of superposition states.

Simultaneous existence of a cat in both death and live states introduced in Schrodinger's thought experiment [1] is impossible in real macroscopic world. Nevertheless it has been realized experimentally [2, 3] through the interaction of atom with coherent state of light as the closest quantum mechanical state to the classical light [4]. These experiments made it possible to study the boundaries of classical and quantum physics [2], and have been adopted to test the basic postulates of quantum mechanics [3]. Quantum nonlocality test [5, 6] and quantum metrology [7, 8] are other essential applications of the superposition states of coherent light.

Many theoretical methods have been proposed to construct quantum light superposition. The Yurke-Stoler state was obtained by the time evolution of coherent light under a time evolution operator based on a nonlinear Hamiltonian describing anharmonic oscillator [9]. Using nonlinear entanglement in the three photon down-conversion process is another method of cat state generation without contribution of the atoms [10]. The even and odd cat states are other known quantum superposition states that are special examples of the Yurke-Stoler states. Dispersive interaction in the Jaynes-Cummings model (JCM) [11], when the atom and photon are largely detuned tends to the even and odd cat states [12]. Powerful Kerr-like nonlinearity [13] is another effect resulting in cat states. Small Kerr nonlinearity is also reported as a method of cat state generation [14]. Preparing the cat states through the interactions in a hybrid system including photons, phonons, and qubit excitations is also theoretically studied in [15].

In this letter, we obtain a general superposition of coherent states based on the exact solution of the JCM. The prominent point of our results is that despite previously introduced methods it is free of any nonlinear effect and don't need any limitation on the interaction parameters. Nevertheless, the general superposition of coherent states converts to many known superposition states and cat states by suitable selection of initial states.

Consider a two level atom by the transition frequency $\omega_A$ interacting with a single mode of optical field of frequency $\omega_F$ for which the interaction intensity is given by $\lambda$. In a system of units for which $\hbar = 1$, this system describes by the Hamiltonian

$$\hat{H} = \frac{\omega_A}{2}\hat{\sigma}_z + \omega_F \hat{a}^\dagger \hat{a} + \lambda(\hat{a}^\dagger + \hat{a})(\hat{\sigma}_+ + \hat{\sigma}_-), \qquad (1)$$

in which $\hat{\sigma}_z$ and $\hat{a}^\dagger \hat{a}$ are the atomic transient operator and the photon number operator, respectively. The field operator is given by $\hat{a}^\dagger + \hat{a}$ and $\hat{\sigma}_+ + \hat{\sigma}_- = \hat{\sigma}_x$ represents the atomic dipole operator. The product $(\hat{a}^\dagger + \hat{a})(\hat{\sigma}_+ + \hat{\sigma}_-)$ models the atomic dipole interaction that consists of the terms $\hat{a}^\dagger \hat{\sigma}_-$, $\hat{a}\hat{\sigma}_+$, $\hat{a}^\dagger \hat{\sigma}_+$ and $\hat{a}\hat{\sigma}_-$. First two terms describe the destruction of an exited atom by creating a photon and exiting an atom by absorbing a photon, respectively while the counter-rotating terms, $\hat{a}\hat{\sigma}_-$ and $\hat{a}^\dagger \hat{\sigma}_+$ describe excitation of an atom simultaneouse with emission of a photon and destruction of an excited atom by absorbing a photon, respectively [16]. The photons included in counter-rotating terms are known as the virtual photons for which the statics and dynamics are studied in [17,18].

We need to obtain the eigenstates of the JCM Hamiltonian given in (1) as the substructures of this letter. Toward this end, we use the convention applied in [19], that is performing a $\pi/2$ Radian rotation of $\hat{H}$ around y axis, $\hat{R}(y, \pi/2)$, to obtain

$$\hat{H}_R = -\frac{\omega_A}{2}\hat{\sigma}_x + \omega_F \hat{a}^\dagger \hat{a} + \lambda(\hat{a}^\dagger + \hat{a})\hat{\sigma}_z. \qquad (2)$$

The rotated Hamiltonian $\hat{H}_R$, commutes with the parity operator, $\hat{\Pi} = \hat{\sigma}_x e^{i\pi\hat{N}}$, so they have the common eigenstates

$$|\psi_{R\pm}\rangle = \tfrac{1}{\sqrt{2}}[|e,\alpha\rangle, \pm|g,-\alpha\rangle], \qquad (3)$$

with the eigenvalues $E_\pm = \mp\tfrac{\omega_A}{2}e^{-2\alpha^2} + \omega_F|\alpha|^2 + \lambda(\alpha + \alpha^*)$. To obtain the eigenstate of the Hamiltonian (1), we just need to operate its eigenstates, $|\psi_{R\pm}\rangle$, with the operator, $\widehat{R}(y, -\pi/2)$ to reverse the rotation. The results are

$$|\psi_\pm\rangle = \tfrac{1}{2}[|e,\alpha\rangle + |g,\alpha\rangle \pm (\ |g,-\alpha\rangle - |e,-\alpha\rangle)]. \qquad (4)$$

Factoring out the atomic states in (4), the eigenstates $|\psi_+\rangle$ and $|\psi_-\rangle$ take the form

$$|\psi_+\rangle = \tfrac{1}{2}[|e\rangle(|\alpha\rangle - |-\alpha\rangle) + |g\rangle(\ |\alpha\rangle + |-\alpha\rangle)], \qquad (5)$$
$$|\psi_-\rangle = \tfrac{1}{2}[|e\rangle(|\alpha\rangle + |-\alpha\rangle) + |g\rangle(\ |\alpha\rangle - |-\alpha\rangle)]. \qquad (6)$$

Defining $|C_e\rangle = |\alpha\rangle + |-\alpha\rangle$ and $|C_o\rangle = |\alpha\rangle - |-\alpha\rangle$ reveals that $|\psi_+\rangle$ and $|\psi_-\rangle$ are combinations of the atomic states with the even and odd cat states, as follows

$$|\psi_+\rangle = \tfrac{1}{2}[|e\rangle|C_o\rangle + |g\rangle|C_e\rangle], \qquad (7)$$
$$|\psi_-\rangle = \tfrac{1}{2}[|e\rangle|C_e\rangle + |g\rangle|C_o\rangle]. \qquad (8)$$

The states $|\psi_+\rangle$ and $|\psi_-\rangle$ given by (7) and (8) are the eigenstates of the Hermitian operator (1), so they compose a complete set and any given state can be expanded as the linear combination

$$|\psi(0)\rangle = c_+|\psi_+\rangle + c_-|\psi_-\rangle. \qquad (9)$$

In which $c_+$ and $c_-$ are complex numbers. Operating with the unitary operator, $\widehat{U} = e^{i\widehat{H}t}$, one can calculate the time evolution of the given state $|\psi(0)\rangle$ as

$$|\psi(t)\rangle = e^{i\widehat{H}t}[c_+|\psi_+\rangle + c_-|\psi_-\rangle]. \qquad (10)$$

From $\widehat{H}|\psi_+\rangle = E_+|\psi_+\rangle$ and $\widehat{H}|\psi_-\rangle = E_-|\psi_-\rangle$, one obtaines

$$|\psi(t)\rangle = c_+ e^{iE_+t}|\psi_+\rangle + c_- e^{iE_-t}|\psi_-\rangle, \qquad (11)$$

that is a general form of Schrodinger's cat state as a result of the atom-photon interaction in the presence of virtual photons. Here we give some examples to simply convert (11) to some known cat states.

As the first example, set $c_+ = c_- = \tfrac{1}{\sqrt{2}}$ in (9) to obtain the initial state

$$|\psi(0)\rangle = \tfrac{1}{\sqrt{2}}(|\psi_+\rangle + |\psi_-\rangle). \qquad (12)$$
Replacing $|\psi_+\rangle$ and $|\psi_-\rangle$ from (4) yields

$$|\psi(0)\rangle = \tfrac{1}{\sqrt{2}}[|e\rangle + |g\rangle]|\alpha\rangle, \qquad (13)$$

that is a normalized initial state in which the atom is initially in the mixed state and the field in the coherent state. It is straightforward to obtain the time evolution of this initial state by using (11) as

$$|\psi(t)\rangle = \tfrac{1}{\sqrt{2}}(e^{iE_+t}|\psi_+\rangle + e^{iE_-t}|\psi_-\rangle). \qquad (14)$$

Some simple calculations after replacing $|\psi_+\rangle$ and $|\psi_-\rangle$ from (4), tends to

$$|\psi(t)\rangle = \tfrac{2A(t)}{\sqrt{2}}(cos\theta(t)[|g\rangle + |e\rangle]|\alpha\rangle - isin\theta(t)[|g\rangle - |e\rangle]|-\alpha\rangle), \qquad (15)$$

in which $A(t) = e^{i(\omega_F|\alpha|^2 + \lambda(\alpha+\alpha^*))t}$ and $\theta(t) = \tfrac{\omega_A t}{2}e^{-2\alpha^2}$. An atomic state measurement in (15) projects the field state into

$$|\psi(t)\rangle = \tfrac{2A(t)}{\sqrt{2}}(cos\theta(t)|\alpha\rangle \pm isin\theta(t)|-\alpha\rangle). \qquad (16)$$

The plus and minus signs appear when the atomic states detected to be in the excited state and the ground state, respectively. The field states $|\alpha\rangle$ and $|-\alpha\rangle$ may be detected with the probabilities $p_+ = cos^2(\tfrac{\omega_A t}{2}e^{-2\alpha^2})$ and $p_- = sin^2(\tfrac{\omega_A t}{2}e^{-2\alpha^2})$ respectively. At $t = 0$, one obtains $p_+ = 1$ and $p_- = 0$ that means the initial state is $|\alpha\rangle$. The probabilities $p_+$ and $p_-$ as functions of α are plotted in the figure 1.

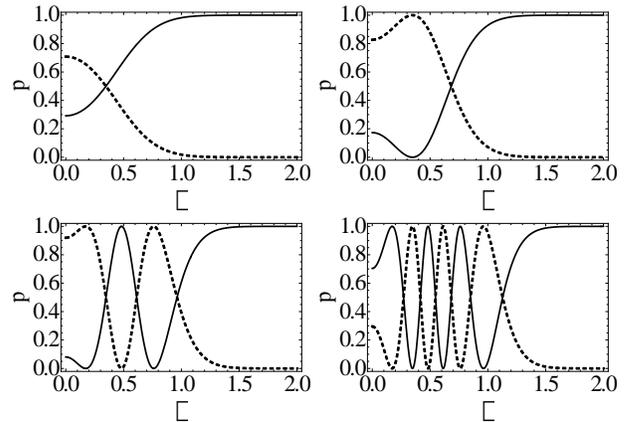

Figure 1. The probabilities of detecting $|\alpha\rangle$ (solid lines) and $|-\alpha\rangle$ (dashed lines) for $\tfrac{\omega_A t}{2} = 1$, top-left, $\tfrac{\omega_A t}{2} = 2$, top-right, $\tfrac{\omega_A t}{2} = 5$, down-left and $\tfrac{\omega_A t}{2} = 10$, down-right, based on (16).

Based on the figure 1, in small times, the probabilities are aperiodic, while for large times, the probabilities are periodic in some intervals of α. As time goes by, the α range for which the probabilities are periodic, increases. For sufficiently large times, the initial probability values will be restored. The probabilities are periodic functions of time which frequencies relate to photon numbers by $e^{-2\alpha^2}$. As shown in the figure 2, increasing α decreases the frequency of the probabilities oscillations in time.

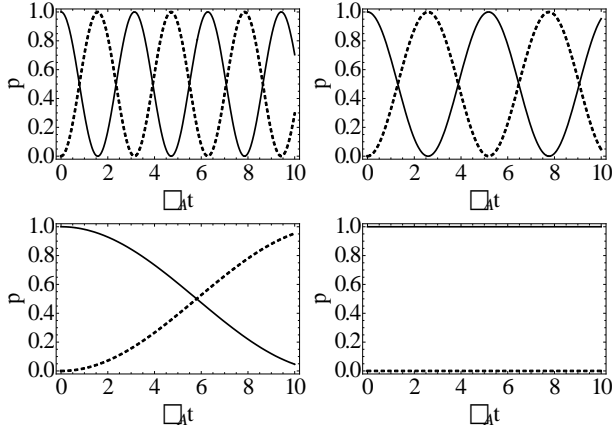

Figure 2. The probabilities of detecting $|\alpha\rangle$ (solid lines) and $|-\alpha\rangle$ (dashed lines) for $\alpha = 0$, top-left, $\alpha = 0.5$, top-right, $\alpha = 1$, down-left and $\alpha = 2$, down-right, based on (16).

For any limited time, there are sufficiently large light intensities for which the initial probabilities will restored, but for any desired large α, there is a time for which the oscillations appear. The light intensity is a parameter by which the frequencies of the probabilities can be controlled to oscillate in time. For $\theta = \pi/2$ the state given by (16) converts to the Yurke-Stoller state

$$|\psi(t)\rangle = A(t)(|\alpha\rangle \pm i|-\alpha\rangle), \qquad (17)$$

as given in [9].

As another example, let $c_+ = \frac{1}{\sqrt{2}}$ and $c_- = -\frac{1}{\sqrt{2}}$ to obtain

$$|\psi(0)\rangle = \frac{1}{\sqrt{2}}(|\psi_+\rangle - |\psi_-\rangle). \qquad (18)$$

That is equivalent to the atom initially in the mixed state and the field in the coherent state, as

$$|\psi(0)\rangle = \frac{1}{\sqrt{2}}[|g\rangle + |e\rangle]|-\alpha\rangle. \qquad (19)$$

For witch the time evolution is

$$|\psi(t)\rangle = \frac{1}{\sqrt{2}}(e^{iE_+t}|\psi_{1+}\rangle - e^{iE_-t}|\psi_{1-}\rangle), \qquad (20)$$

based on the equation (11). Some simple calculations after replacing $|\psi_+\rangle$ and $|\psi_-\rangle$ from (4), tends to

$$|\psi(t)\rangle = \frac{2A(t)}{\sqrt{2}}(-i\sin\theta(t)[|g\rangle + |e\rangle]|\alpha\rangle + \cos\theta(t)[|g\rangle - |e\rangle]|-\alpha\rangle). \qquad (21)$$

An atomic state measurement in (21) projects the field state into

$$|\psi(t)\rangle = \frac{2A(t)}{\sqrt{2}}(-i\sin\theta(t)|\alpha\rangle \mp \cos\theta(t)|-\alpha\rangle). \qquad (22)$$

The minus sign appears when excited atom detects and the plus sign appears when detected atom is measured in ground state. The field states $|\alpha\rangle$ and $|-\alpha\rangle$ may be detected with the probabilities $p_+ = sin^2(\frac{\omega_A t}{2}e^{-2\alpha^2})$ and $p_- = cos^2(\frac{\omega_A t}{2}e^{-2\alpha^2})$ respectively. For $\frac{\omega_A t}{2}=0$ the probabilities are $p_+ = 0$ and $p_- = 1$. The probabilities change with $\frac{\omega_A t}{2}$ for $\alpha < 1$ and for large values of α, the initial probability values will be restored. The results are similar to those of (19) but the probabilities $p_+$ and $p_-$ are interchanged. For $\theta = \pi/2$ the state given by (22) converts to

$$|\psi(t)\rangle = A(t)(-i|\alpha\rangle \mp |-\alpha\rangle). \qquad (23)$$

That is perpendicular to the Yurke-Stoler state (17).

Another interesting example can be demonstrated by $c_+ = 1$ and $c_- = 0$ to obtain the initial state

$$|\psi(0)\rangle = |\psi_+\rangle. \qquad (24)$$

That is equivalent to $\frac{1}{\sqrt{2}}[|e\rangle|C_o\rangle + |g\rangle|C_e\rangle]$ that evolves by time according to

$$|\psi(t)\rangle = e^{iE_+t}|\psi_+\rangle. \qquad (25)$$

Replacing $|\psi_+\rangle$ from (4) gives

$$|\psi(t)\rangle = \frac{1}{\sqrt{2}}e^{iE_+t}[|e\rangle|C_o\rangle + |g\rangle|C_e\rangle], \qquad (26)$$

for which, the atomic state measurement switches the light state to the even or odd cat states with equal probabilities if the measured atomic state appears to be in ground or excited states, respectively.

Here we propose an experimental setup to construct desired cat state based on the initial atomic and field states. As demonstrated schematically in the figure 3, the atoms prepared in the oven, rotate around $y$ axis as much as $\pi/2$ Radian by passing the region R, then the atoms lead to a cavity to interact with the coherent light emitted by the field source. Affecting by an inverse rotator, $R^{-1}$, is the next step after which the quantum state of the system will be a linear combination of $|\psi_+\rangle$ and $|\psi_-\rangle$ which coefficients related to the initial atomic and field states. At last an atomic state detector used to switch the light state to some light state superposition. As described in different examples, the output state of light depends on the initial atomic and light states. The setup given in the figure 3 creates not only the even and odd cat states given in [10], but also the atom independent superposition states of coherent light reported in [9], so it can be considered as general cat state generator.

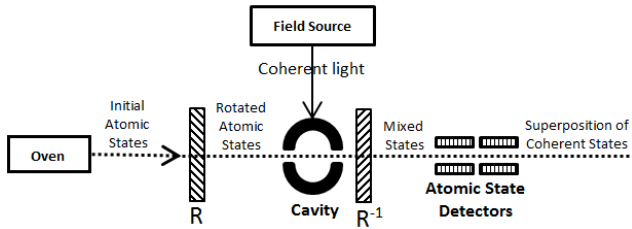

Figure 3. The experimental setup based on the proposed method for cat state generation.

As demonstrated in the examples, our approach tends to a general superposition of coherent states by taking into account the role of virtual photons in the JCM. The even and odd cat states can be obtained by special selection of the initial states while the previous method needs a dispersive atom-photon interaction before entering the second Ramsey zone.

Another striking point of this work is that it generates the well-known Yurke-Stoler states by suitable selection of initial state in the same procedure demonstrated in the figure 3 while it is reported based on the time evolution of the coherent state under a nonlinear Hamiltonian governing an anharmonic oscillator [10].

Finally we note that the virtual photons play the role of nonlinear effects in the Yurke-Stoler cat states and the role of the dispersive interaction in the even and odd cat states and unifies the famous cat states in a simple interaction. It is another valuable point that these states are simple examples of the calculated general cat state and it is possible to extract some unknown cat states.

**References**
   * m-mirzaee@araku.ac.ir

[1] E. Schrödinger, *Naturwissenschaften,* 23: pp.807-812; 823-828; 844-849 (1935). Translated in J. A. Wheeler, W. H. Zurek, Eds., Quantum Theory and measurement, Princeton. Univ. Press, Princeton, NJ, (1983).
[2] C. Monroe, D. M. Meekhof, B. E. King, D. J. Wineland, Science, 272, (1996).
[3] S. Haroch, M. Brune, J. M. Raimond, Phil. Trans. R. Soc. Lond. A 355, 2367-2380 (1997).
[4] D. F. Walls, Gerard J. Milburn, Quantum optics, Springer, (2008).
[5] D. Wilson, H. Jeong and M. S. Kim, J. Mod. Opt. 49, 851 (2002).
[6] H. Jeong et al., Phys. Rev. A 67, 012106 (2003).
[7] T. C. Ralph, Phys. Rev. A 65, 042313 (2002).
[8] W. J. Munro et al ., Phys. Rev. A 66, 023819 (2002).
[9] B. Yurke, D. Stoler, Phys. Rev. Lett. 57, (1986).
[10] Y. Shen, S. M. Assad, N. B. Grosse, X. Y. Li, M. D. Reid, and P. K. Lam, Phys. Rev. Lett. 114, 100403 (2015).
[11] E.T. Jaynes and F. W. Cummings, "Comparison of quantum and semiclassicalradiation theories with application to the beam mase," Proc.IEEE, 51(1) 89, (1963).
[12] C. C. Gerry, P. L. Knight, Am. J. Phys 66, 10 (1997).
[13] C. M. Savage, S. L. Braunstein, D. F. Walls, Opt. Lett. 15, 11 (1990).
[14] H. Jeong, M. S. Kim, T. C. Ralph, and B. S. Ham, Phys. Rev. A 70, 061801(R) (2004).
[15] Mehdi Abdi, Matthias Pernpeintner, Rudolf Gross, Hans Huebl, and Michael J. Hartmann, Phys. Rev. Lett. **114**, 173602 (2015).
[16] Wolfgang P. Schleich, Quantum optics in phase space, Wiley-VCH, Berlin (2001).
[17] C. K. Law, Phys. Rev. A 87, 045804 (2013)
[18] G Compagnot, G M Palmat, R Passantet and F Persicott, J. phys. B At. Mol. Opt. Phys., 28 1105 (1995).
[19] M. Mirzaee and M. Batavani, *Chin. Phys. B* **24,** 040306 (1995).